# An Overview of Rossby Wave Instability in Accretion Discs surrounding Black Holes

B. B. Dutta[1], L. Devi[2], B. Sarkar[2], A. J. Boruah[2,*]

[1]Department of Mechanical Engineering, Tezpur University, Tezpur - 784028, Assam, India

[2]Department of Applied Sciences, Tezpur University, Tezpur - 784028, Assam, India

*Corresponding author, Email: app23109@tezu.ac.in

**Abstract.**

The Rossby Wave Instability (RWI) has become an important concept in understanding the hydrodynamics (HDs) of accretion discs (ADs), especially in systems around black holes (BHs) where magnetic effects are either weak or absent. This instability is triggered by extrema (or sharp gradients) in the vortensity profile of the disc. Once activated, it leads to non-axisymmetric disturbances that can grow into large-scale vortices. These vortices play a significant role in the outward transport of angular momentum (AM). They may also help explain the presence of quasi-periodic oscillations (QPOs) observed in certain astrophysical systems such as X-ray binaries (XRBs). Here we review the main theoretical ideas behind RWI, as well as findings from more advanced three-dimensional (3D) and relativistic simulations. We also mention how the theory has been extended to include magnetic fields and self-gravity (SG) and what these results might imply for actual observations.

## 1 Introduction

In the extreme environments surrounding BHs, ADs act as engines of energy release, driven in part by complex mechanisms of AM redistribution. In ionized discs, where the magnetic field is weak, magneto-rotational instability (MRI) provides a magnetic channel, while HD alternatives, including RWI, become important in weakly ionized thin discs than in the partially or fully ionized discs [1, 2, 3]. The results of Gholipour & Nejad-Asghar [4] confirmed this.

To describe large-scale atmospheric motions on rotating planets, Rossby waves (RWs) were originally introduced and because of conservation of potential vorticity (PV)[#] in rotating fluids, RWs arise [5, 6]. They play a fundamental role in shaping large-scale flow patterns. In planetary atmospheres, they are responsible for the meandering of jet streams, the formation of cyclones and anticyclones, and long-term weather variability. These waves occur when there is a variation in the Coriolis parameter with latitude. This leads to the propagation of large-scale, low-frequency disturbances. Hence, their study is crucial to atmospheric dynamics and planetary-scale circulation [3, 4]. Over time, researchers recognized that the underlying physics of RWs wasn't limited to planetary atmospheres. Lovelace et al. [1] and Li et al. [8] developed RWI for thin ADs with minimal SG after Lovelace & Hohlfeld [9] initially proposed RWI for thin disc galaxies. In such discs, sharp gradients or extrema in a quantity called entropy-modified vortensity can trigger what's known as the RWI. When this happens, instead of fading away like typical linear disturbances, large, coherent vortices can form. Due to their high-pressure nature, these vortices tend to accumulate and retain dust particles. They are robust, nonlinear structures that not only survive over time but also play an active role in reshaping the disc. These waves can transport AM outward, influence accretion processes, and even contribute to the early stages of planet formation in protoplanetary discs (PPDs) [1, 2, 6, 8]. In order to attain intriguing results in AM transport and planetary formation through thin ADs, RWI theory has emerged in recent years both through analytical studies [7, 8, 9, 10] and numerical simulations [11, 12, 13, 14].

---

[#]The PV is proportional to the dot product of stratification (stratification means the gradient of entropy) and vorticity. It indicates the amount to which the rotation of the flow aligns in connection with the entropy gradient (from reference [5]).

This review examines the evolution of understanding of RWI in the context of ADs around BHs, tracing its origins in linear theory, its nonlinear development into vortex structures, and its extention to 3D and relativistic effects. It also examines how magnetic fields and SG influence RWI and considers its possible connection to observed QPOs in BH systems.

## 2   Linear and Nonlinear Evolution of RWI

Lovelace et al. [1] and Li et al. [8] constructed a comprehensive linear theory of this non-axisymmetric instability and discovered that the RWI can be triggered when an extremum exists in the radial profile of an entropy-modified form of PV. The RWI arises in differentially rotating discs when there exists an extremum in a key function $L(r)$. $L(r)$ is defined as:

$L(r) \equiv F(r) S^{\frac{2}{\Gamma}}(r)$, where $F^{-1} = \frac{\hat{z}.(\nabla \times v)}{\Sigma}$, $v$ is the flow velocity of the disc, $S = \frac{P}{\Sigma^{\Gamma}}$ is the specific entropy, $\Sigma$ is the surface density, and $\Gamma$ is the adiabatic index [1].

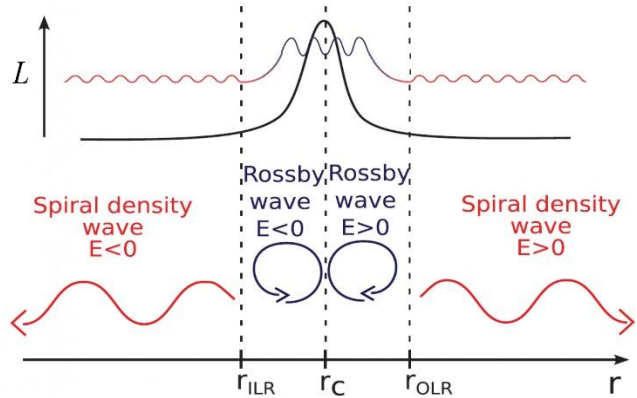

Figure 1: Diagrammatic representation of the RWI showing the two zones where density waves and RWs propagate, as well as the region between the evanescent zones (from Fig. 1 of Meheut et al. [15]).

There can be two standing RWs in a disc with an extremum, one on each side of the extremum (Fig. 1) [15]. The wave in the positive gradient of $L$ ($r < r_c$) has a frequency pattern that is less than the AD's rotation frequency and contains negative energy. The wave in the negative gradient of $L$, on the other hand, conveys positive energy and has a frequency greater than the disc. The interaction between the two RWs, which have positive and negative energy, respectively, causes the RWI to grow. As the wave amplitude rises, the overall energy is preserved [15]. Then, independent of the boundary conditions (BCs), the mechanism for the instability's development is localized in the co-rotation zone. RWs and density waves are coupled in ADs by the differential rotation. These density waves only propagate outside of the Lindblad resonance [Lindblad resonances are termed as ideal absorbers of density waves, which travel in a primary range between the inner and outer Lindblad resonances (ILR and OLR, respectively) [16, 17]] radii ($r_{ILR}$ & $r_{OLR}$). As seen in Fig. 1, the wave is evanescent between the two propagation areas, and tunneling is plausible [15].

As RWI grows and moves beyond its initial stage, it reaches a point of nonlinear saturation, where its behaviour becomes more complex. Li et al. [2] tested the nonlinear development of the instability using in-depth global two-dimensional (2D) HD simulations. They have demonstrated that vortices naturally form in the nonlinear stage, which involves a region of high pressure and density. These vortices are counter spinning, or anticyclonic, yet they are (almost) corotating with the background flow. They have demonstrated that these vortices are long-lived structures inside discs and have explained the mechanism responsible for their formation, which is the azimuthal pressure gradients. It is demonstrated that these vortices play a crucial role in both developing global accretion and the outward transport of AM [2]. They discovered that the nonlinear consequences of the instability become less pronounced when the cooling period is short. The efficiency of any transport processes will be severely limited if the cooling time is less than one rotation cycle, because changes brought on by the vortex motion will probably be damped out very rapidly [2].

Meheut [13] provides an overview of some earlier findings regarding the nonlinear evolution, saturation, development, and structural formation of RWI in a stratified disc, considering both radial and vertical gradients. According to Meheut [13], RWI can increase linearly in PPDs if a process can create an extremum of a particular vorticity. Depending on its thermodynamics, the vertical stratification can introduce a vertical component into the flow's structure, but it does not significantly alter the instability [13].

## 3    3D and Relativistic Effects

Initial studies of RWI considered 2D models of ADs, but more realistic models require a 3D approach, as actual accretion flows exhibit vertical stratification and complex dynamics in the radial, azimuthal, and vertical directions. Meheut et al. [11] were the first to introduce RWI's full simulations in 3D for stratified cylindrical discs. These simulations validated the basic expectations of earlier 2D predictions of vortex formation. A powerful vortex that expands quickly and reaches maximum amplitude is visible in the simulation. However, the third dimension exhibits surprising extra characteristics that may be crucial to understanding the astrophysical functions of these vortices [11].

In relativistic ADs, where the effects of strong gravitational fields near BHs become significant, the dynamics of RWI exhibit increased complexity. Vincent et al. [14] extended previous studies by using ray-traced simulations and considering relativistic corrections. They have examined a model that focuses on the presence of RWI at the disc's inner edge. In fact, it was demonstrated that the disc is prone to RWI when a BH has an innermost stable circular orbit (ISCO) around it because of a natural extremum in the vorticity profile. They demonstrated that the flow can be modulated by the 3D RWI while being within the observed range.

## 4    Magnetic Field Effects

Although in classical HDs, the RWI is often discussed in the context of unmagnetized discs, it has been found that it remains active in magnetized discs under specific conditions. Through linear analysis, Yu & Li [18] demonstrated that the RWI in a disc may be totally reduced by a weak magnetic field which is toroidal. At the corotation resonance, the magnetic field modifies wave absorption, which leads to a reduction of RWI. Again, Yu & Lai [19] investigated the behaviour of RWI in ADs threaded by large-scale poloidal magnetic fields. Under sub-thermal (plasma $\beta \sim 10$, $\beta$ being the ratio of thermal pressure to magnetic pressure) conditions for RWI, the magnetic field can still have an impact. The instability can be increased by roughly 10% for discs that are infinitely thin. As the field becomes closer to equipartition ($\beta \sim 1$), the growth rate of RWI increases significantly (by a factor of ~2) in a disc with finite thickness and radial magnetic field intensity gradient.

How toroidal magnetic fields influence RWI's non-axisymmetric behaviour in the self-gravitating AD was also studied by Gholipour & Nejad-Asghar [4]. According to their findings, the toroidal magnetic field and viscosity will both dampen RW perturbations, whereas magnetic diffusivity works the other way around. The occurrence of RWI is also dependent on the turbulent magnetic Prandtl number.

Using the magnetohydrodynamics (MHD) code PLUTO, the RWI in a narrow AD connected by an originally toroidal magnetic field is investigated by Matilsky et al. [20]. They discovered that the late time and the growth rate density are decreased for plasma, where $10^1 < \beta < 10^3$, while the magnetic field has very little effect for plasma having $\beta > 10^3$.

## 5 RWI with SG

In non-self-gravitating discs, such as those surrounding recently formed stars, an axisymmetric bump in the disc surface mass-density at a certain radius $r_0$ may cause instability. Around $r_0$, it results in an exponentially rising non-axisymmetric perturbation made up of anticyclonic vortices [10].

Lovelace & Hohlfeld [21] have examined the instability of RWs in both the self-gravitating thin AD model and the non-self-gravitating thin AD model. The Toomre parameter (Q), the azimuthal wave number (which has no dimension) of the perturbation, and the parameters of the bump (or depression) in the inverse PV (i.e. $F(r)$) at $r_0$ are the key factors determining the stability or instability. With sound speed ($c_s$) and a certain radius ($r_0$), the value of Q is proportional to $\frac{\dot{M}}{\alpha}$ ($\alpha$ being the viscosity having no dimension and $\dot{M}$ is the accretion rate (AR)), which determines the regime of instability. Their findings imply that SG should be included in Rossby vortex simulations when $Q < \frac{r_0}{h}$, where $h$ is the disc's half-thickness.

Khosravi & Khesali [22] investigated how various viscosity models affected Toomre instability and RWs in self-gravitating ADs. According to their findings, viscosity can amplify Toomre instability in ADs, which suggests that Toomre instability could be a promising option for planetesimal formation. In systems with low Q, SG might generate turbulence, and this turbulence's viscosity is dependent on Q. Consequently, the viscosity of systems may be increased when SG is present. RWI modes were found to nearly vanish at the high limit of $\alpha$ (= 0.2) when the azimuthal mode number is small ($m = 1$). The findings indicate that in strong SG AD with moderate or high viscosity, RWI cannot actively contribute to planetesimal development for small azimuthal mode numbers [22]. According to the study by Su & Wei [23], the cooling parameter ($\Lambda$) and the disc's SG indicated by mass-averaged Q-parameter ($\bar{Q}$) significantly influence the form of gravitational instability (GI) in PPDs. The author demonstrated how GI activates the transitional modes for effective AM transport, disc substructures (such as spirals and fragmentations), and brown dwarf/planet formation in the outer disc at $Q \approx 1$ and $\Lambda \approx 1$.

BH ADs are basically different from PPDs. The disc's SG has a direct effect on PPDs. When the $Q \lesssim 2$, RWI can develop at grooves or bumps in the surface density, creating anticyclonic vortices that support the development of planetesimals [21]. On the other hand, in most cases, the mass of the disc is typically regarded as negligible in comparison to the mass of the compact object sitting in the center of a BH system [24]. Their SG is therefore irrelevant. Here, general relativistic (GR) impacts have a major impact on the RWI's growth rates and saturation rate (explained in section 3) [14].

# 6 Observational Signatures

One of the most intriguing and astrophysical relevant implications of RWI lies in its potential to explain high-frequency QPOs (HFQPOs) observed in BH XRBs and their origin [25]. Expanding on this idea, Tagger & Varniere [26] proposed a hybrid model that combines the dynamics of the accretion-ejection instability (AEI) with those of the MHD variant of RWI. They demonstrated that RWI instability caused by circumstances close to a BH disc's marginally stable orbit (MSO) can account for the HFQPOs seen in a small number of microquasars. This instability can be characterized as a combination of the AEI and a closely related but distinct instability, which is an MHD form of the RWI. Using data from RXTE's XTE J1550−564, Varniere et al. [25] found that a disc with an active RWI falls in the same fitted parameter space as XRB systems where an HFQPO is found. Additionally, the authors have extended the relationship between the HFQPOs and appearance of the RWI in the inner region of the AD by demonstrating a clear distinction between RWI-activated discs and systems that exhibit a type-C low-frequency QPO with missing HFQPOs.

Tagger & Melia [27] also have shown that the RWI might account for the quasi-periodic behaviour detected during the galaxy's central BH flares in Sgr A*, which occurs when a blob of ambient gas is trapped in the disc at a few to a few tens of Schwarzschild radius ($R_s$).

GRHD simulations of 2D discs swirling around rotating BHs and subject to the RWI were carried out by Casse & Varniere [28]. The RWI may form regardless of the spin parameter and at any point on the disc, according to them. They discovered that when RWI develops in the extremely close vicinity of BHs having high spin where GR effects are dominant, the saturation threshold of instability rises considerably.

RWI (or RWI-like instabilities) in GRHD/GRMHD ADs around rotating BHs have been linked to possible gravitational wave (GW) emission in both mathematical and real-world simulations. With cooling creating the most noticeable effect on final GW emission, Gottlieb et al. [29] discovered an overall pattern across simulations, making this the most promising case. They claim that collapsar ADs may be the most viable burst-type GW sources so far, based on the comparatively high estimated event rate. Current detectors may be able to detect the GWs in the near future.

## 6.1 Laboratory Evidence for HD Transport

Most studies of RWI lean heavily on simulations and analytical work, but real-world experiments paint a convincing picture of purely HD transport. In quasi-Keplerian regimes, the turbulent viscosity parameter is on the order of $10^{-5}$, according to experimental data from Taylor–Couette (TC) flows that were originally examined by Richard & Zahn [30] and then integrated by Dubrulle et al. [31]. Both theoretical and observational research on ADs, especially in fully ionized regimes, usually conclude that the Shakura–Sunyaev (SS) viscosity parameter, α, falls between about 0.1 and 0.4 [32]. Thus, the laboratory-derived value of the viscosity parameter is many orders of magnitude less than the α-values for astrophysical discs, although operating in quite different physical regimes. This implies that whereas TC experiments capture basic features of HD turbulence, AM transport in ADs is probably enhanced by other mechanisms (such as MRI and radiative processes).

Under simplified but well-founded assumptions, Hersant et al. [33] suggested that the equations driving an α-disc may be described in the TC flow and are comparable to the governing equation of motion of an incompressible shear flow having rotation with penetrable BCs and cylindrical configuration.

The scaling of $\frac{G}{Re^2}$ (*G* is dimensionless torque and *Re* is Reynolds' number) is proportional to both the $\dot{M}$ and turbulent viscosity parameter, making it the most astrophysically relevant conclusion. Paoletti et al. [34] discovered that the estimated $\dot{M}$ for quasi-Keplerian flows is around 14% of the highest rate possible for a given *Re*. The author calculated the value of AM transport parameter as $2 \times 10^{-5}$, which is consistent with Richard and Zahn [30]'s findings. AM transport can yield dimensionless ARs on the order of $\frac{\dot{M}}{\dot{M}_\odot} \sim 10^{-3}$, which is consistent with observations in discs surrounding T Tauri stars [33].

When two concentric cylinders revolving at differing angular velocities destabilize the Newtonian fluid's flow, this is known as TC instability. When viscous dissipation is driven out by the centrifugal force's destabilizing impact, this instability arises [35]. The kinematic inertia force is not taken into account by Taylor's instability criteria. It just examines the impact of centrifugal force. As a result, it may be suitable for low *Re* numbers having high curvature. A rotating flow with a low curvature and a higher *Re* number can shift to turbulence faster without breaking Taylor's instability criteria [36]. This limitation mainly separates TC flow when compared with real astrophysical systems.

# 7     Conclusion

RWI plays an important role in settling disputes between MRI and HD in dead zones (DZs). Gammie [37] originally suggested that there is a DZ in the AD of Young Stellar Objects (YSO). The author discovered that cosmic rays ionize the outer regions of the disc while collisions ionize the inner region. However, there is a middle zone, known as DZ, where the ionization is so low that the gas is not magnetically coupled. This is problematic since it is generally accepted that MHD instabilities, like the MRI, cause accretion [38]. According to Varnière & Tagger [39], who reshaped the view of DZ, the existence of the DZ self-consistently produces a density profile that is similar to the RWI [1]. Rossby vortices will develop and continue to exist in the disc as a result of this instability and might encourage the creation of planets. Additionally, they demonstrated how this instability propagates across DZ as spiral waves, which may sustain a significant AR even in the absence of locally generated turbulence.

The ultimate purpose of the upcoming investigations will be to better understand the theoretical models utilizing precise GRMHD simulations. In order to validate the power spectra that were observed, researchers will examine the parameters that help the growth of many azimuthal modes by RWI, along with how they interplay [40, 25, 13]. With the development of astrophysical machine learning (ML) tools, researchers may use ML to study the QPO frequency generated by RWI. Researchers then need to train both supervised and unsupervised ML models utilizing both synthetic observational data and raw simulation data. Researchers might find and validate realistic RWI behavior in the extreme conditions around BH by combining developed ML models with observable astronomical data obtained by detectors like the Atacama Large Millimeter/submillimeter Array (ALMA)/Event Horizon Telescope (EHT). With the construction of the ALMA telescope, which enables the first mapping of the distribution of grains of dust and gas present in the ADs of neighboring protoplanetary systems, the study on vortices in astrophysical discs is currently under development [10]. The combination of simulation, theory, and observation should strengthen RWI's position in high-energy astrophysics.

# References

[1] R. V. E. Lovelace, H. Li, S. A. Colgate, and A. F. Nelson, "*Rossby wave instability of Keplerian accretion disks*," Astrophys. J., **vol. 513, no. 2**, p. 805, 1999.

[2] H. Li, S. A. Colgate, B. Wendroff, and R. Liska, "*Rossby Wave Instability of Thin Accretion Disks. III. NonlinearSimulations*," Astrophys. J., **vol. 551, no. 2**, p. 874, 2001.

[3] M. Gholipour, "*Rossby wave instability in the accretion flows around black holes*," Astrophys. J., **vol. 835, no. 1**, p. 18, 2017.

[4] M. Gholipour and M. Nejad-Asghar, "*The role of toroidal magnetic field on the Rossby wave instability*," Mon. Not. R. Astron. Soc., **vol. 449, no. 2**, pp. 2167–2173, 2015.

[5] P. G. Drazin, "*Rotating fluids in geophysics*." London: Academic Press, 1978.

[6] S. Helmreich, *A book of waves*. Duke University Press, 2023.

[7] T. V. Zaqarashvili *et al.*, "*Rossby Waves in Astrophysics*," Space Sci. Rev., **vol. 217, no. 1**, p. 15, Feb. 2021, doi: 10.1007/s11214-021-00790-2.

[8] H. Li, J. M. Finn, R. V. E. Lovelace, and S. A. Colgate, "*Rossby Wave Instability of Thin Accretion Disks. II. DetailedLinear Theory*," Astrophys. J., **vol. 533, no. 2**, p. 1023, 2000.

[9] R. V. E. Lovelace and R. G. Hohlfeld, "*Negative mass instability of flat galaxies*," Astrophys. J., **vol. 221**, pp. 51–61, 1978.

[10] R. V. E. Lovelace and M. M. Romanova, "*Rossby wave instability in astrophysical discs*," Fluid Dyn. Res., **vol. 46, no. 4**, p. 041401, 2014.

[11] H. Meheut, F. Casse, P. Varniere, and M. Tagger, "*Rossby wave instability and three-dimensional vortices in accretion disks*," Astron. Astrophys., **vol. 516**, p. A31, 2010.

[12] H. Meheut, R. Keppens, F. Casse, and W. Benz, "*Formation and long-term evolution of 3D vortices in protoplanetary discs*," Astron. Astrophys., **vol. 542**, p. A9, 2012.

[13] H. Meheut, "*The Rossby wave instability in protoplanetary disks*," in EPJ Web of Conferences, EDP Sciences, 2013, p. 03001.

[14] F. H. Vincent, H. Meheut, P. Varniere, and T. Paumard, "*Flux modulation from the Rossby wave instability in microquasars' accretion disks: toward a HFQPO model*," Astron. Astrophys., **vol. 551**, p. A54, 2013.

[15] H. Meheut, R. V. E. Lovelace, and D. Lai, "*How strong are the Rossby vortices?*," Mon. Not. R. Astron. Soc., **vol. 430, no. 3**, pp. 1988–1993, 2013.

[16] E. V. Polyachenko and I. G. Shukhman, "*Effect of inner Lindblad resonance on spiral density waves propagation in disc galaxies: reflection over absorption*," Mon. Not. R. Astron. Soc., **vol. 483, no. 1**, pp. 692–703, 2019.

[17] C. C. Lin and F. H. Shu, "*On the Spiral Structure of Disk Galaxies, II. Outline of A Theory of Density Waves*," Proc. Natl. Acad. Sci., **vol. 55, no. 2**, pp. 229–234, Feb. 1966, doi: 10.1073/pnas.55.2.229.

[18] C. Yu and H. Li, "*Nonaxisymmetric Rossby Vortex Instability with Toroidal Magnetic Fields in Radially Structured Disks*," Astrophys. J., **vol. 702, no. 1**, p. 75, 2009.

[19] C. Yu and D. Lai, "*Rossby wave instability in accretion discs with large-scale poloidal magnetic fields*," Mon. Not. R. Astron. Soc., **vol. 429, no. 3**, pp. 2748–2754, 2013.

[20] L. Matilsky, S. Dyda, R. V. E. Lovelace, and P. S. Lii, "*Rossby vortices in thin magnetized accretion discs*," Mon. Not. R. Astron. Soc., **vol. 480, no. 3**, pp. 3671–3679, 2018.

[21] R. V. E. Lovelace and R. G. Hohlfeld, "*Rossby wave instability with self-gravity*," Mon. Not. R. Astron. Soc., **vol. 429, no. 1**, pp. 529–533, 2013.

[22] A. Khosravi and A. Khesali, "*The effect of different models of viscosity on the Rossby wave and Toomre instability in accretion disks with self gravity*," Astrophys. Space Sci., **vol. 361, no. 1**, p. 13, Jan. 2016, doi: 10.1007/s10509-015-2602-2.

[23] Z. Su and X. Wei, "*Gravitational instability in protoplanetary disk with cooling: 2D global analysis*," Astrophys. J., **vol. 983, no. 2**, p. 89, 2025.

[24] G. Lodato, "*Self-gravitating accretion discs*," Riv. Nuovo Cimento, no. Online First, pp. 293–99999, Dec. 2007, doi: 10.1393/ncr/i2007-10022-x.

[25] P. Varniere, F. Casse, and F. H. Vincent, "*Rossby wave instability and high-frequency quasi-periodic oscillations in accretion discs orbiting around black holes*," Astron. Astrophys., **vol. 625**, p. A116, 2019.

[26] M. Tagger and P. Varniere, "*Accretion-ejection instability, MHD Rossby wave instability, diskoseismology, and the high-frequency QPOs of microquasars*," Astrophys. J., **vol. 652, no. 2**, p. 1457, 2006.

[27] M. Tagger and F. Melia, "*A possible Rossby wave instability origin for the flares in Sagittarius A*," Astrophys. J., **vol. 636, no. 1**, p. L33, 2005.

[28] F. Casse and P. Varniere, "*On the Rossby wave instability in accretion discs surrounding spinning black holes,*" Mon. Not. R. Astron. Soc., **vol. 481, no. 2**, pp. 2736–2744, 2018.

[29] O. Gottlieb, A. Levinson, and Y. Levin, "*In ligo's sight? vigorous coherent gravitational waves from cooled collapsar disks*," Astrophys. J. Lett., **vol. 972, no. 1**, p. L4, 2024.

[30] D. Richard and J.-P. Zahn, "*Turbulence in differentially rotating flows What can be learned from the Couette-Taylor experiment*," Astronomy and Astrophysics, **vol. 347**, pp. 734–738, 1999.

[31] B. Dubrulle, O. Dauchot, F. Daviaud, P.-Y. Longaretti, D. Richard, and J.-P. Zahn, "*Stability and turbulent transport in Taylor–Couette flow from analysis of experimental data*," Phys. Fluids, **vol. 17, no. 9**, 2005.

[32] A. R. King, J. E. Pringle, and M. Livio, "*Accretion disc viscosity: how big is alpha?*," Mon. Not. R. Astron. Soc., **vol. 376, no. 4**, pp. 1740–1746, 2007.

[33] F. Hersant, B. Dubrulle, and J.-M. Huré, "*Turbulence in circumstellar disks*," Astron. Astrophys., **vol. 429, no. 2**, pp. 531–542, 2005.

[34] M. S. Paoletti, D. P. van Gils, B. Dubrulle, C. Sun, D. Lohse, and D. P. Lathrop, "*Angular momentum transport and turbulence in laboratory models of Keplerian flows*," Astron. Astrophys., **vol. 547**, p. A64, 2012.

[35] A. Meyer, I. Mutabazi, and H. N. Yoshikawa, "*Stability of Rayleigh-stable Couette flow between two differentially heated cylinders*," Phys. Rev. Fluids, **vol. 6, no. 3**, p. 033905, Mar. 2021, doi: 10.1103/PhysRevFluids.6.033905.

[36] H.-S. Dou, B. C. Khoo, and K. S. Yeo, "*Instability of Taylor–Couette flow between concentric rotating cylinders*," Int. J. Therm. Sci., **vol. 47, no. 11**, pp. 1422–1435, 2008.

[37] C. F. Gammie, "*Layered accretion in T Tauri disks*," Astrophys. J., **vol. 457**, p. 355, 1996.

[38] S. A. Balbus and J. F. Hawley, "*A powerful local shear instability in weakly magnetized disks. I-Linear analysis. II-Nonlinear evolution*," Astrophys. J., **vol. 376**, pp. 214–233, 1991.

[39] P. Varnière and M. Tagger, "*Reviving Dead Zones in accretion disks by Rossby vortices at their boundaries*," Astron. Astrophys., **vol. 446, no. 2**, pp. L13–L16, 2006.

[40] C. Cui, A. Tripathi, C. Yu, M.-K. Lin, and A. Youdin, "*Rossby wave instability in magnetized protoplanetary discs–I. Azimuthal or vertical B-fields*," Mon. Not. R. Astron. Soc., **vol. 537, no. 2**, pp. 1973–1983, 2025.